\documentclass[%
reprint,
superscriptaddress,
 amsmath,amssymb,
 aps,
]{revtex4-2}
\usepackage[utf8]{inputenc}
\usepackage[T1]{fontenc}
\usepackage{gensymb}
\usepackage{physics}
\usepackage{graphicx}
\usepackage{dcolumn}
\usepackage{bm}

\begin{document}

\title{Ultra-bright entangled-photon pair generation from an AlGaAs-on-insulator microring resonator}
\author{Trevor J. Steiner}
\altaffiliation[these authors ]{contributed equally to this work}
\affiliation{Materials Department, University of California, Santa Barbara, CA 93106}

\author{Joshua E. Castro}
\altaffiliation[these authors ]{contributed equally to this work}
\affiliation{Electrical and Computer Engineering Department, University of California, Santa Barbara, CA 93106}

\author{Lin Chang}
\altaffiliation[these authors ]{contributed equally to this work}
\affiliation{Electrical and Computer Engineering Department, University of California, Santa Barbara, CA 93106}

\author{Quynh Dang}
\affiliation{Electrical and Computer Engineering Department, University of California, Santa Barbara, CA 93106}
\author{Weiqiang Xie}
\affiliation{Electrical and Computer Engineering Department, University of California, Santa Barbara, CA 93106}
\author{Justin Norman}
\affiliation{Electrical and Computer Engineering Department, University of California, Santa Barbara, CA 93106}
\author{John E. Bowers}
\affiliation{Materials Department, University of California, Santa Barbara, CA 93106}
\affiliation{Electrical and Computer Engineering Department, University of California, Santa Barbara, CA 93106}

\author{Galan Moody}
\email{moody@ucsb.edu}
\affiliation{Electrical and Computer Engineering Department, University of California, Santa Barbara, CA 93106}

\date{\today}


\begin{abstract}
Entangled-photon pairs are an essential resource for quantum information technologies. Chip-scale sources of entangled pairs have been integrated with various photonic platforms, including silicon, nitrides, indium phosphide, and lithium niobate, but each has fundamental limitations that restrict the photon-pair brightness and quality, including weak optical nonlinearity or high waveguide loss. Here, we demonstrate a novel, ultra-low-loss AlGaAs-on-insulator platform capable of generating time-energy entangled photons in a $Q$ $>1$ million microring resonator with nearly 1,000-fold improvement in brightness compared to existing sources. The waveguide-integrated source exhibits an internal generation rate greater than 20$\times10^9$ pairs sec$^{-1}$ mW$^{-2}$, emits near 1550 nm, produces heralded single photons with $>99\%$ purity, and violates Bell’s inequality by more than 40 standard deviations with visibility $>97\%$. Combined with the high optical nonlinearity and optical gain of AlGaAs for active component integration, these are all essential features for a scalable quantum photonic platform.

\end{abstract}
\maketitle

\maketitle
\section{Introduction}

Entanglement is the cornerstone of quantum science and technologies. Compared to matter-based quantum systems, such as electronic spins \cite{Togan2010}, optomechanical resonators \cite{Riedinger2018}, superconducting circuits \cite{Houck2012}, and trapped atoms and ions \cite{Moehring2007}, photons are unique in their ability to generate and distribute entangled quantum states across long distances in free space or fiber networks whilst retaining a high degree of coherence. Following the seminal experiments demonstrating the creation of photon pairs in maximally entangled Bell states over two decades ago, parametric down conversion in bulk nonlinear crystals has been the workhorse for quantum light generation \cite{Kwiat1995NewHS}. Despite the low efficiency of bulk sources, they produce nearly indistinguishable photons with high purity and entanglement fidelity at nearly gigahertz pair generation rates, making them appealing for a wide range of experiments, including foundational tests of quantum mechanics \cite{Shalm2015,Giustina2015}, sensing \cite{Pirandola2018} , information processing \cite{Flamini2019}, and satellite-based quantum communications \cite{Yin2017,Liao2017}.

Recent advances in the fabrication of quantum photonic integrated circuits (QPICs) have enabled the functionality of benchtop nonlinear sources to be scaled down to a single chip with dramatic improvement in the efficiency and stability \cite{Wang2019,Elshaari2020,Wang2020,Moody2020}. Silicon-based photonics is a versatile platform for QPICs owing to its relative low waveguide loss and existing foundry infrastructure developed for the telecommunications industry. State-of-the-art integrated quantum photonic circuits based on SOI are capable of implementing quantum gates between two qubits \cite{Qiang2018} and chip-to-chip teleportation \cite{Llewellyn2019}, for example, but they have to rely on off-chip detectors that introduce significant loss, slow thermal-based active components for tuning and modulation, and off-chip high-power lasers to generate single and entangled photons due to the moderate optical nonlinearity of silicon. Other nonlinear material platforms have been developed to address some of these issues, including lithium niobate \cite{Zhao2020}, aluminum nitride \cite{Guo2017}, indium phosphide \cite{Kumar2019}, silicon carbide\cite{Guidry2020}, and heterogeneous approaches \cite{Kim2020,Elshaari2020}, but each has tradeoffs including higher waveguide loss and complex fabrication that hinders the scalability and reliability. Moreover, it is not readily apparent that any of these platforms are capable of all-on-chip integration of lasers, sources, active and passive components, and detectors---all which are essential for creating practical and scalable QPIC technologies \cite{Kim2020}. 

\begin{figure*}[tb]
\centering
\includegraphics[width=6.21in, height=3.68in]{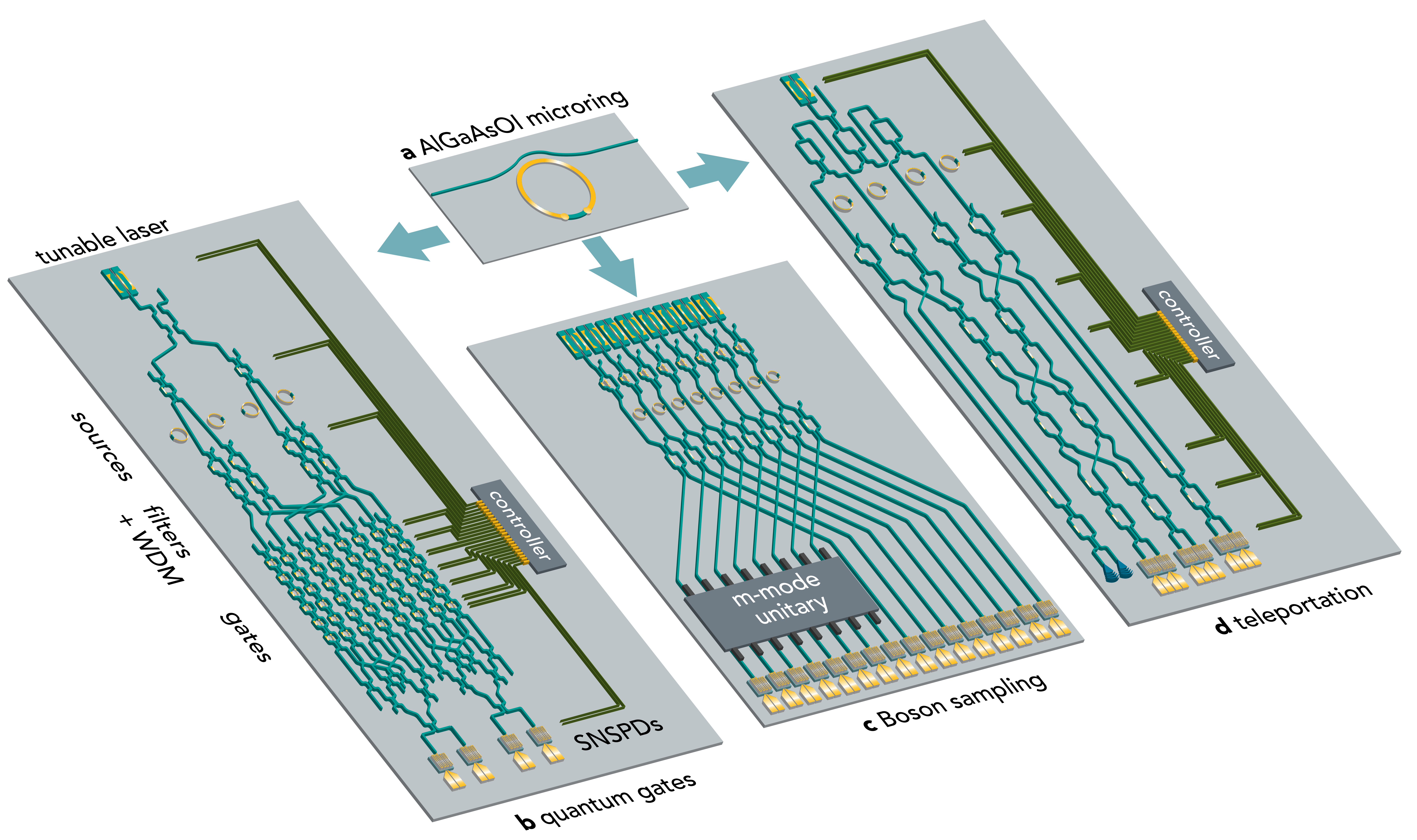}
\caption{(a) Tunable AlGaAs-on-insulator (AlGaAsOI) microring resonator entangled-photon pair source. The AlGaAsOI platform enables the large-scale integration of active and passive quantum photonic components, including tunable lasers, nonlinear quantum light sources, filters and wavelength division multiplexing (WDM), superconducting nanowire single-photon detectors (SNSPDs), and microcontrollers. These components can be monolithically integrated for all-on-chip quantum photonic circuits, including (b) quantum gates for optical computing, (c) \textit{m}-mode unitary operations for Boson sampling, and (d) Bell-state measurements for chip-to-chip teleportation of quantum states.}
\label{fig:devices}
\end{figure*}

Here, we report a new QPIC platform based on AlGaAs-on-insulator (AlGaAsOI) that has exciting prospects for all-on-chip quantum information applications. Compound semiconductors have been widely used in the photonics industry due to their attractive properties of light emission and well-developed material growth and fabrication processes. They also hold great potential for nonlinear processes since these materials usually exhibit orders of magnitude higher nonlinear coefficients compared to dielectrics. However, previous compound semiconductor photonic platforms suffer from weak optical confinement on native substrates and significant waveguide loss that has limited their utility for quantum photonic applications. Over the past few years, a novel platform, AlGaAsOI, has emerged by heterogeneously integrating (Al)GaAs onto an oxidized Si substrate that is capable of providing high index contrast waveguides \cite{Pu2016}.  More importantly, recent breakthroughs in the processing techniques on this platform have significantly reduced the propagation loss for AlGaAsOI waveguides. Currently, for fully etched submicron-scale AlGaAsOI waveguides, the loss is $<0.2$ dB/cm, resulting in microring resonator quality factors $Q>3\times10^6$, over-performing the SOI platform. Such high-$Q$ resonators have enabled a record-low threshold frequency comb generations at 1550 nm \cite{Chang2020,Xie2020}, under a pump power at the level of tens of micro-Watts, which opens an unprecedented highly nonlinear regime for integrated photonics.

In this work, we make a key step to bring the desired nonlinear properties of this platform into the quantum regime through spontaneous four-wave mixing, where pump photons tuned into resonance with a comb line near 1550 nm can be annihilated to produce time-energy entangled signal and idler photon pairs at adjacent comb lines \cite{Helt2010}. We demonstrate that the combination of high $Q$, small microring radius, large $\chi^{(3)}$ nonlinearity, and tight modal confinement factor of AlGaAsOI results in more than a 500-fold improvement in the time-energy entangled-pair generation rate $PGR = 20\times10^9$ pairs sec$^{-1}$ mW$^{-2}$ over state-of-the-art with $97.1\pm0.6\%$ visibility, $>4300$ coincidence-to-accidental ratio (CAR), and heralded single-photon antibunching $g_H^{(2)}(0)<0.004\pm0.01$.

In addition to the entangled-pair source demonstrated here, AlGaAsOI has remarkable potential for all-on-chip QPIC development compared to existing platforms. First, the on-chip integration of tunable excitation laser sources can be naturally incorporated into the epitaxial growth process of the AlGaAs photonic layer \cite{Liu2018}. AlGaAsOI is also distinguished by high index contrast for tight modal confinement \cite{Ottaviano2016}, has negligible two-photon absorption at 1550 nm with proper Al portion \cite{Adachi1994}, exhibits a large $\chi^{(2)}$ nonlinearity for high-speed electro-optic modulation \cite{Walker2019}, strong piezo-optic effect for optomechanic cavities \cite{Forsch2020}  and ultra-quiet superconducting nanowire single-photon detectors (SNSPDs) have been integrated with GaAs/AlGaAs waveguides \cite{McDonald2019}. The possible application space is extraordinarily broad (Figure \ref{fig:devices}), ranging from ground-to-satellite communications and quantum teleportation to all-on-chip quantum information processing and Boson sampling \cite{Yin2017,Liao2017,Llewellyn2019,Spring2012}.

\section{Methods}

\subsection{Device fabrication}
In this work, the AlGaAs photonic layer was grown by Molecular-Beam Epitaxy (MBE). Its layer structure from top to substrate is: a [001] orientated 400-nm-thick Al$_{0.2}$Ga$_{0.8}$As film on a 500-nm-thick Al$_{0.8}$Ga$_{0.2}$As layer on a GaAs substrate. A 5-nm-thick Al$_2$O$_3$ film was deposited on the epi-layer by Atomic Layer Depositon (ALD) as an adhesive layer for bonding. The wafer was then bonded on to a 3-$\mu$m-thick thermal SiO$_2$ buffer layer on a Si substrate. The thermal SiO$_2$ layer was pre-patterned by Inductively Coupled Plasma (ICP) etch. The surfaces of both chips were treated by atmospheric plasma before bonding to activate the surface. After initial contact, the bonded sample was placed in an oven at 100 $^{\circ}$C for 12 hours under 1 MPa pressure to enhance the bonding strength.

Removal of the GaAs substrate was performed in three steps. First, mechanical polishing was applied to lap the GaAs substrate down to ~70 $\mu$m. Then the remaining substrate was removed by H$_2$O$_2$:NH$_4$OH (30:1) wet etch. Finally, the Al$_{0.8}$Ga$_{0.2}$As buffer layer was selectively etched by diluted hydrofluoric (~2.5\%) acid, leaving only the Al$_{0.2}$Ga$_{0.8}$As photonic layer on the carrier wafer.

After substrate removal, a 5-nm ALD Al$_2$O$_3$ layer was deposited on Al$_{0.2}$Ga$_{0.8}$As for surface passivation, followed by a 100-nm SiO$_2$ layer deposition as a hardmask. The wafer was then patterned by a deep ultraviolet (DUV) stepper using a photoresist (UV6$^{TM}$-0.8). Prior to the photoresist coating, an anti-reflective (AR) coating (DUV-42P) was used to suppress the backreflection during photolithography. After exposure and development of the resist, a thermal reflow process was applied to the wafer at 155 $^{\circ}$C on a hotplate for 2 minutes. ICP etches using O$_2$ and CHF$_3$/CF$_4$/O$_2$ gases were used to remove the AR coating and define the hardmask, respectively, followed by another ICP etching using Cl$_2$/N$_2$ gases to pattern the Al$_{0.2}$Ga$_{0.8}$As layer. After the etch, the sample was passivated by 5-nm Al$_2$O$_3$ layer by ALD and finally clad with 1.5-$\mu$m thick SiO$_2$ by PECVD. 

\subsection{Singles and coincidences measurements}
In order to measure the singles rate for the entangled photon pairs generated from the AlGaAsOI microring resonator, a continuous wave laser was swept to and held at the ring resonance wavelength. After attenuating the laser power to achieve the desired on-chip power and sending the laser through fiber-based bandpass filters to suppress amplified spontaneous emission (ASE) at the signal and idler wavelengths, the light was coupled onto and off of the AlGaAsOI chip via an inverse taper waveguide. Light couples to the microresonator via a pulley waveguide design. The coupling loss was approximately 5 dB per facet. The output power from the resonator was then sent through a fiber Bragg grating (FBG) and split using a 50/50 fiber-based beamsplitter into the signal and idler channels. For the signal channel, the pump and idler photons were filtered by over 100 dB using a series of fiber-based tunable etalon filters. The idler channel used a similar cascade that was tuned to filter out the pump and signal photons. After the filter array, the signal and idler channels were coupled to superconducting nanowire single-photon detectors (SNSPDs) from PhotonSpot to detect the photons on each channel. Using a time-correlated single-photon counting (TCSPC) module, the average signal and idler counts were determined over 10 minutes integration. An example scan is included in Appendix \ref{app:singlescan}. The scan was started with the laser set to a slightly off-resonance wavelength to determine the background counts on the detectors (from both the dark counts of the system and any pump photons that reach the detector). 

For the coincidence measurements, the TCSPC module was set to trigger when a photon was detected on the signal channel and record the time-delay until a photon arrived on the idler channel. An example coincidence measurement is shown in Appendix \ref{app:coinc}. The coincidence measurements were recorded simultaneously with the singles rates to assure consistency with the reported values of on-chip power and loss in the system. For the lower power measurements, the integration time for the coincidence measurements was extended to allow for the calculation of the CAR. Since at low powers, the overall number of photons and the background counts on the detectors was very low, the average number of accidentals was often zero for short integration times. Thus, the resonator was pumped at resonance for integration times up to 180 minutes until the average accidental counts minus the standard deviation were greater than zero. A complete diagram and discussion of the experimental design can be found Appendix \ref{app:exp}.

\subsection{Franson interferometry and visibility measurements} 
To assess the two-photon interference visibility of the entangled photon pairs, a folded Franson interferometric setup was added to the experimental design between the FBG and the chip. The schematic shown in Figure \ref{fig:hist}a demonstrates the experimental setup with the interferometer. In an identical fashion to the singles and coincidences measurements, the laser was swept and held at the ring resonance wavelength after going through a variable optical attenuator (VOA) and ASE suppression filters. The output from the resonator is coupled through a 50/50 fiber-based beamsplitter into a short and a long path with a path length difference of approximately 7 meters. A fiber phase shifter was placed on the short arm to allow for the modification of the phase of the photons. The reported 7 meter path difference includes the length added to the long arm by the phase shifter. The phase shifter was locked to the pump power using the drop port output from the FBG and a servo loop filter with a phase stability better than $\lambda/100$. Coincidence measurements were completed for 10 minute integration times with the phase stepped from 0.4$\pi$ to 1.4$\pi$. As the phase approaches $\pi$, the number of coincidences at zero time delay approach zero. 

\subsection{Heralded g$^{(2)}$(0) measurements}

Heralded g$^{(2)}_H$ measurements are performed via three-fold coincidence detection with a third detector. The signal photons are sent to one PhotonSpot SNSPD as the herald. The idler photons are sent into a 3 dB fiber beam splitter with the outputs connected to the other PhotonSpot SNSPD and a single-photon avalanche diode (SPAD) from ID Quantique. Using the TCSPC module, we record singles counts from the heralding detector ($N_A$), coincidence counts between the heralding detector and each of idler detectors ($N_{AB}$ and $N_{AC}$), and the three-fold coincidence counts $N_{ABC}$. A coincidence window of 4 ns is used as determined from the two-fold coincidence histogram width. The timing between the channels was calibrated with separate coincidence measurements and the delays adjusted accordingly. Within the total integration time of 300 minutes, we do not observe any three-fold coincidence counts above the background; thus g$^{(2)}_H\left(0\right)$ is estimated from the raw counts (\textit{i.e.} without background subtraction) through the expression $\frac{N_{ABC}N_{A}}{N_{AB}N_{AC}}$.

\section{Results}

\subsection{SFWM from AlGaAsOI Microring Resonators}
\begin{figure*}[ht]
\centering
\includegraphics[width=5.7in, height=4in]{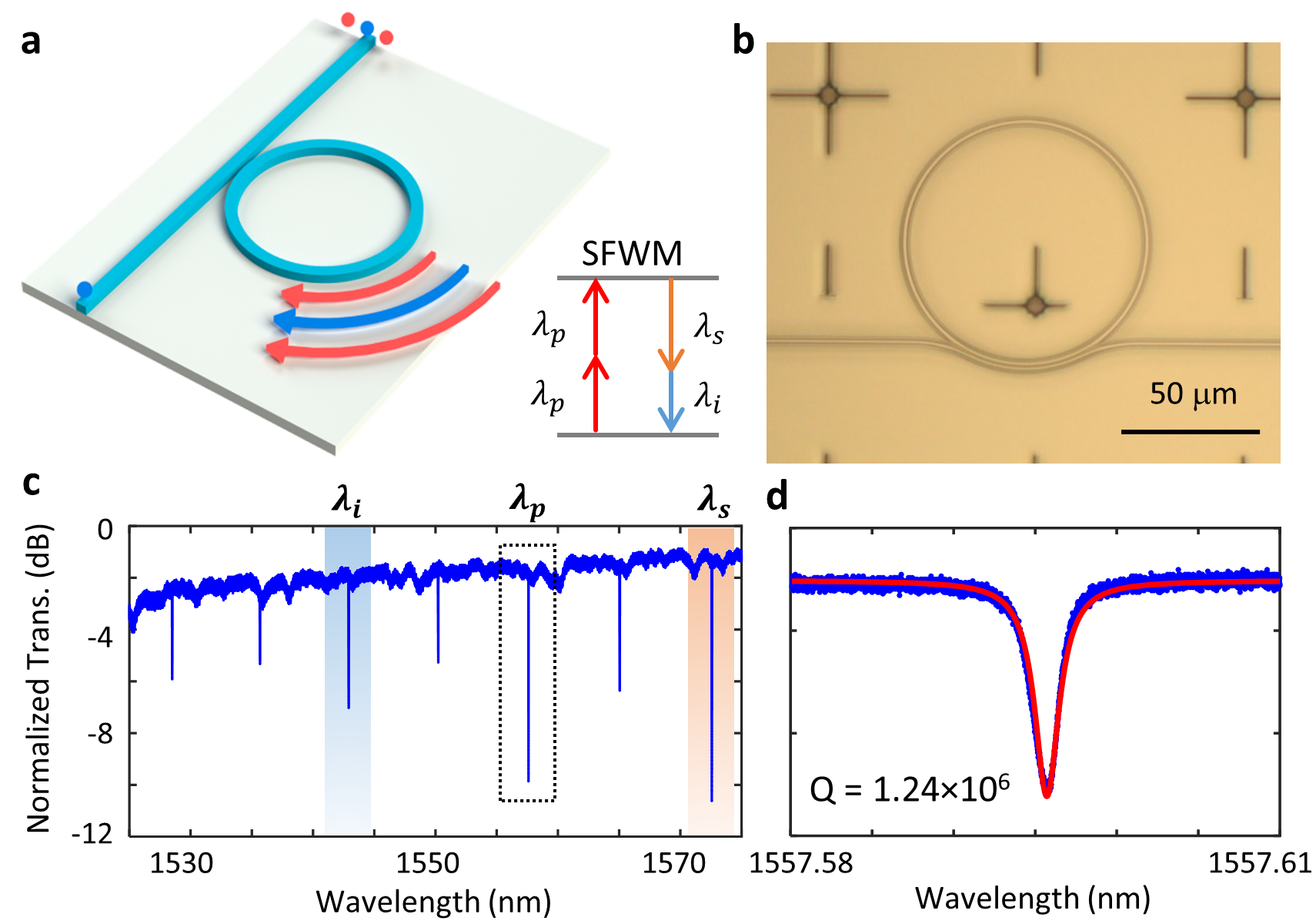}
\caption{(a) Schematic illustration of entangled-photon pair generation from a photonic microring resonator. Pump photons ($\lambda_p$) are coupled into the waveguide bus and ring and are converted into signal ($\lambda_s$) and idler ($\lambda_i$) photons via spontaneous four-wave mixing (SFWM). (b) Optical microscope image of a representative AlGaAsOI ring resonator pulley with 30 $\mu$m radius. (c) Resonator transmission spectrum with signal (1572 nm) and idler (1542 nm) wavelengths two free-spectral ranges away from the pump (1557 nm) resonance. (d) Resonator transmission spectrum of the pump resonance (blue-trace). A $Q = 1.24\times10^6$ is determined from the superimposed Lorentzian fit (red-trace).}
\label{fig:SFWM}
\end{figure*}

The process of spontaneous four-wave mixing (SFWM) in a microring resonator is illustrated in Figure \ref{fig:SFWM}a. The inset depicts the fundamental concept behind SFWM, where two pump photons (denoted $\lambda_p$) are annihilated and a signal photon ($\lambda_s$) and an idler photon ($\lambda_i$) are created. This process only occurs at the resonances of the microring resonator in which quasi-phase matching between the pump, signal, and idler is attained. A demonstrative microscope image of a 30 $\mu$m-radius AlGaAsOI microring resonator and pulley waveguide is shown in Figure \ref{fig:SFWM}b, and the microfabrication procedure is described in more detail in the methods section. The microring resonator studied in the following experiment had a radius of 13.91 $\mu$m. The width of the bus waveguide was 0.48 $\mu$m, and the ring waveguide was 0.69 $\mu$m wide. The gap between the waveguide and the ring was 0.48 $\mu$m, and the AlGaAs layer was 0.4 $\mu$m thick. The transmission spectrum of the ring resonator is shown in Figure \ref{fig:SFWM}c. The sharp troughs indicate resonance wavelengths of the microring resonator separated by the free spectral range (FSR). A high-resolution sweep of the pump comb line is shown by the blue trace in Figure \ref{fig:SFWM}d. The resonance is fit with a Lorentzian function to determine the quality factor \textit{$Q$} of the cavity, which is is proportional to the ratio of the full-width at half-maximum (FWHM) of the transmission resonance to the FSR. In comparison to other microring resonators utilized for entangled-pair generation, the $Q=1.24\times10^6$ measured for this device is a factor of 100 larger than InP, a factor of $\sim$10 larger than AlN and silicon-on-insulator (SOI), and is comparable to Si$_3$N$_4$, as shown in Table 1.

By pumping the resonator at one of these resonance wavelengths, entangled-photon pairs are generated spontaneously at adjacent resonances through SFWM. In the experiments presented here, and shown by the dashed box in Figure \ref{fig:SFWM}c, the pump wavelength is set to be resonant with 1557.59 nm, as this resonance peak had the highest quality factor for the selected ring. As we show in the next section, the on-chip pair generation rate depends on several factors and is given by \cite{Helt2010,Azzini2012,Kumar2019}

\begin{equation}
    \label{Eq:1}
    PGR=\left(\gamma 2\pi R\right)^2 \left(\frac{Q v_g}{\pi\omega_p R}\right)^3 \frac{v_g}{4\pi R} P_p^2,
\end{equation}
where $R$ is the ring resonator radius, $\gamma$ is the nonlinear coefficient of the material that also takes into account the confinement factor, $Q$ is the quality factor, $\omega_p$ is the angular frequency, $v_g$ is the group velocity at the pump wavelength, and $P_p$ is the on-chip pump power. The quality factor $Q$ is a critical parameter for the photon-pair generation rate as it is related to the amount of time that resonant photons reside within the cavity and thus the amount of interaction time photons have with each other. With the pump laser set to this resonance wavelength, entangled-photon pairs are generated at adjacent multiples of the cavity FSR. In our experiments, the second nearest neighbor resonances highlighted in blue and orange in Figure \ref{fig:SFWM}c are selected as the signal and idler photons, since this provided additional pump rejection through the bandpass optical filters before the single-photon detectors. Using measured and calculated properties of our AlGaAsOI microresonators, we expected a $PGR = 10^{10}$ pairs sec$^{-1}$ mW$^{-2}$.

\subsection{On-Chip Photon-Pair Generation Rate and Brightness}
\begin{figure*}[t]
\centering
\includegraphics[width=6.5in, height=2.15in]{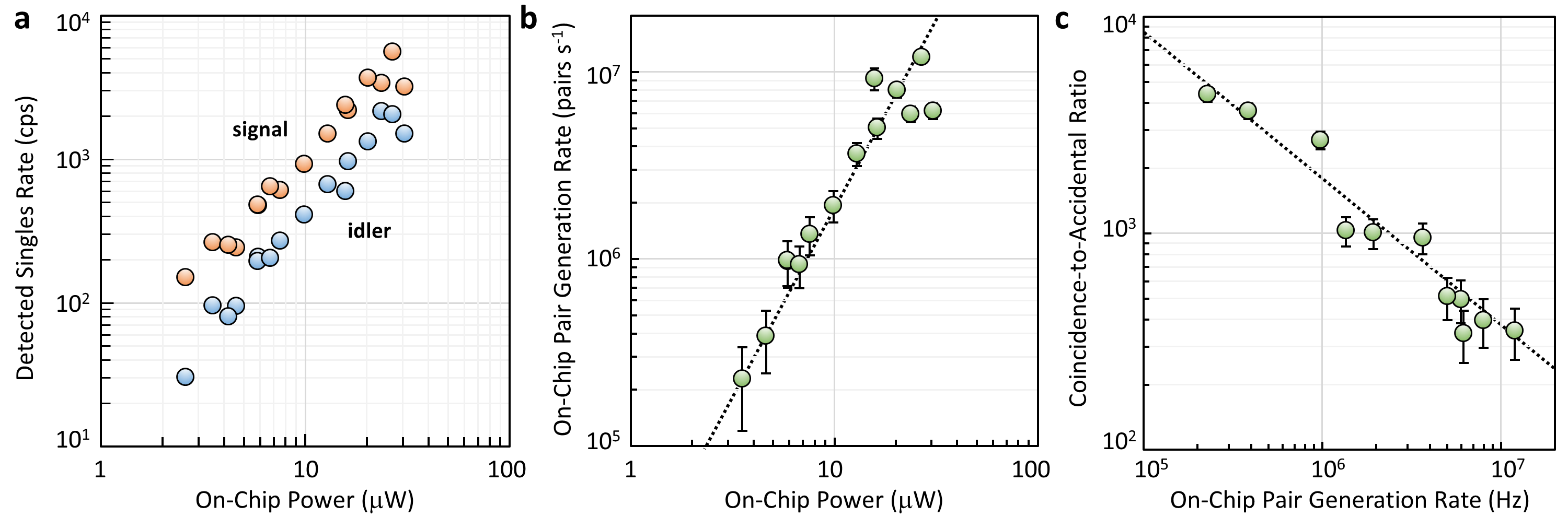}
\caption{(a) Corrected detected singles rates versus on-chip pump power. The difference in the generated singles rates is due to the difference in the filter loss in the two channels (19.4 dB for the idler and 13.6 dB for the signal). (b) On-chip pair generation rate versus on-chip pump power. The dashed line is a fit to the data, yielding a pair generation rate of $20\times10^9$ pairs sec$^{-1}$ mW$^{-2}$ (c) Coincidence-to-accidental ratio (CAR) versus on-chip pair generation rate.}
\label{fig:PGR}
\end{figure*}

A schematic of the fiber-based experimental setup for entangled-pair generation is shown in Appendix \ref{app:exp} and discussed in detail in the Methods section. Briefly, tapered fibers are used to couple the pump laser and entangled pairs onto and off the AlGaAsOI chip. After the chip, the residual pump, signal, and idler photons pass through a 3 dB splitter, and the pump light is filtered using a series of tunable Fabry-Perot etalons to provide more than 130 dB of rejection in each of the signal and idler paths. The signal and idler photons are detected with superconducting nanowire single-photon detectors (SNSPDs) with sub-40 ps timing jitter. The singles rates are shown in Figure \ref{fig:PGR}a for the signal and idler channels for various on-chip pump powers. It is clear that the rates follow a $P_p^2$ dependence as expected for the SFWM process. We note that these values are the raw values as detected at the SNSPDs. The idler filter path has a larger loss (19.4 dB) compared to the signal filters (13.6 dB) at their respective wavelengths, which explains the disparity between the singles rates. In addition to the filter losses, the generated signal and idler rates are also reduced due to the chip-to-fiber coupling loss and loss in the SNSPD system. 

We next measure the coincidence counts by recording two-photon correlation histograms using a time-correlated single-photon counting module. From these measurements, the on-chip PGR is determined by dividing the time-averaged value of the measured coincidence counts, $N_c$, by the total loss from chip to the SNSPDs. These quantities were measured by separate calibration procedures. The on-chip PGR for various on-chip pump power is shown in Figure \ref{fig:PGR}b. Unlike InP and silicon-based microresonators, we do not observe any saturation of the curve due to two-photon absorption (TPA) as expected, since the AlGaAs bandgap wavelength is shorter than 775 nm. The black dashed line fitted to the data illustrates the $P^2_p$ behavior as expected. From this fit we determine a slope of $20\times10^9$ pairs sec$^{-1}$ mW$^{-2}$. To the best of our knowledge, this value is over 100-times higher than any previously reported PGR to date, as shown in Table 1. By normalizing the PGR to the FWHM of the emission resonance ($\sim1$ pm), we obtained the entangled-pair brightness $B=2\times 10^{11}$ pairs sec$^{-1}$ GHz$^{-1}$ (normalized to 1 mW on-chip power), which is more than a 500-fold improvement upon previous state-of-the-art microresonators based on Si$_3$N$_4$ \cite{Ramelow2015} and more than 1000-times brighter than SOI \cite{Ma2017}.

\subsection{Coincidence-to-Accidentals Ratio}
Another important metric of entangled-photon pair sources is the coincidence-to-accidentals ratio (CAR). An example histogram utilized to determine the CAR is shown in Supplementary Information Figure 2. The CAR versus the PGR is shown in Figure \ref{fig:PGR}c. The CAR is calculated as the FWHM of the signal-idler coincidence histogram measured as a function of the inter-channel delay divided by the background counts across a similar time window. The highest CAR we measure is 4389 when the PGR is $2.3\times10^5$ pairs sec$^{-1}$. At the highest PGR measured here ($12\times10^6$ pairs sec$^{-1}$), the CAR is equal to 353. The CAR decreases with increasing pump power as $P^{-1}_p$ as expected and shown by the dashed line in Figure \ref{fig:PGR}c. We were not able to measure higher CAR at lower on-chip pump power, since the majority of the histogram time bins had zero registered background counts from our SNSPDs at these powers even for integration times up to 1.5 hours. For comparison, CAR values for an on-chip $PGR$ = $10^6$ pairs sec$^{-1}$ are shown in Table 1 for various photonic entangled-pair sources. Our reported value is a factor of 4 larger than the next highest reported value at this PGR (LiNbO$_3$ periodically poled waveguide).

\subsection{Time-Energy Entanglement}
\begin{figure*}[ht]
\centering
\includegraphics[width=6.5in, height=2.6in]{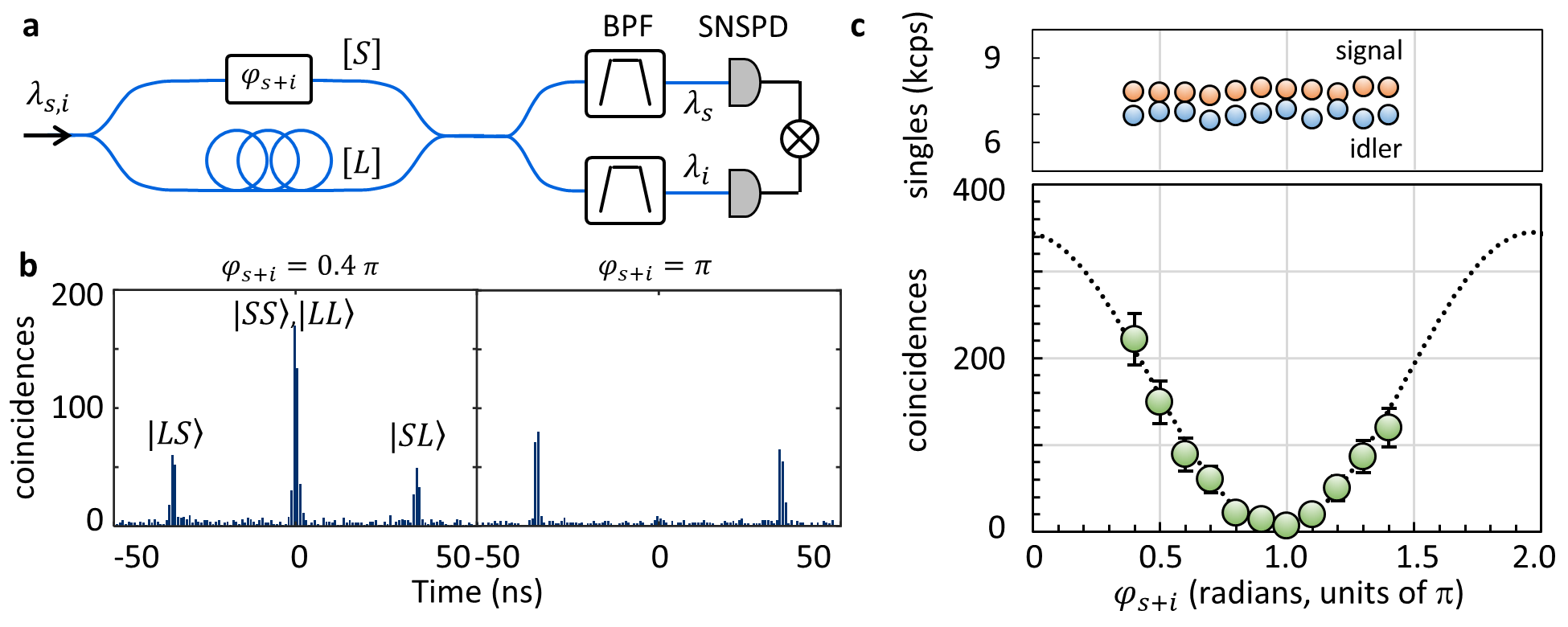}
\caption{(a) Schematic illustration of the setup for the two-photon interference experiment using a fiber-based folded Franson interferometer. BPF = bandpass filter; [S] ([L]) denotes the short (long) interferometer path; SNSPD = superconducting nanowire single-photon detector. (b) Coincidence histograms for different interferometric phase $\phi_{s+i}$ = 0.4$\pi$ and $\pi$, respectively.(c) Singles counts (top) measured with the interferometer setup simultaneously with the two-photon interference (bottom). Visibility of the raw (fitted) data yields $95\%$ ($97.1\%$).}
\label{fig:hist}
\end{figure*}

The generated signal-idler pair is expected to exhibit time-energy entanglement \cite{Brendel1999,Wakabayashi2015}, which can be measured through a Franson-type two-photon interference experiment as depicted in Figure \ref{fig:hist}a \cite{Franson1989,Franson1991}. Signal and idler photons travel through the unbalanced Mach-Zehnder interferometer and then are separated by a 3 dB splitter and bandpass filters before arriving at the SNSPDs. The interferometer path length difference is set such that the propagation delay $\Delta$t is longer than the single-photon coherence time $\tau_c$ in order to avoid single-photon interference at the detectors, but shorter than the laser coherence time $\tau_L$. In this case, the signal and idler photons can travel along either the short [S] or long [L] paths, allowing a total of four possible permutations. 

The differences in photon arrival times between the two paths are illustrated in Figure \ref{fig:hist}b. We can express the two-photon state as a summation over $\ket{ij}$, where the $i$ ($j$) index is the path the signal (idler) photon travels, with $i,j$ = [S,L]; however, post-selection allows for the different states to be distinguished. The side peaks arise from photons travelling along the $\ket{LS}$ or $\ket{SL}$ paths and are offset from zero delay by $\Delta t$. The central peak at zero delay is due to both photons taking the same paths, $\ket{SS}$ or $\ket{LL}$. Because these states are indistinguishable, the two-photon state is expressed as $\frac{1}{\sqrt{2}}\left(\ket{SS} + exp(i\phi_{i+s})\ket{LL}\right)$. By inserting a voltage-controlled fiber phase shifter into the short path and sweeping the phase, two-photon interference is observed as shown in Figure \ref{fig:hist}b for two different phases $\varphi_{s+i} = 0.4\pi$ and $\pi$. The coincidence counts versus phase is shown in the bottom panel of Figure \ref{fig:hist}c. Proof of photon entanglement requires the Clauser-Horne interference pattern visibility $V\geq70.7\%$ \cite{Clauser1974,Rarity1990}. Calculated from the raw data (fitted data), we obtain $V = 95\%$ ($V = 97.1\%$), measured when the on-chip PGR was approximately $10^6$ pairs sec$^{-1}$. These measurements confirm the high quality of the time-energy entangled-pair source, as illustrated by the variation of the two-photon coincidences and the constant signal and idler singles rates versus applied phase shift indicating the absence of single-photon interference.

\subsection{Heralded Single-Photon Generation}

Detection of one of the photons from the pair projects the other photon into a multi-mode single-photon state that is expected to exhibit non-classical anti-bunching. The heralded (\textit{i.e.} conditional) single-photon second-order auto-correlation function, $g_H^{\left(2\right)}\left(0\right)$, is measured by detecting one photon from the pair and performing an auto-correlation measurement on the remaining photon using a Hanbury Brown and Twiss type interferometric experiment. We calculate the auto-correlation function from $g_H^{\left(2\right)}\left(0\right)$ = $\frac{N_{ABC}N_{A}}{N_{AB}N_{AC}}$, where $N_A$ is the average singles counts on the heralding detector, $N_{AB}$ and $N_{AC}$ are the coincidence counts between the heralding and either of the other detectors, and $N_{ABC}$ is the three-fold coincidence counts between all detectors. We calibrate the arrival times of pulses from all detectors using separate measurements, and we average counts within a 4 ns second window centered at zero delay. We measure $g_H^{\left(2\right)}\left(0\right) = 0.004\pm0.01$ for an on-chip pump power corresponding to a PGR approximately equal to $10^6$ pairs sec$^{-1}$. 

\section{Conclusion}

\begin{figure}[ht]
\centering
\includegraphics[width=\columnwidth, height=3.3in]{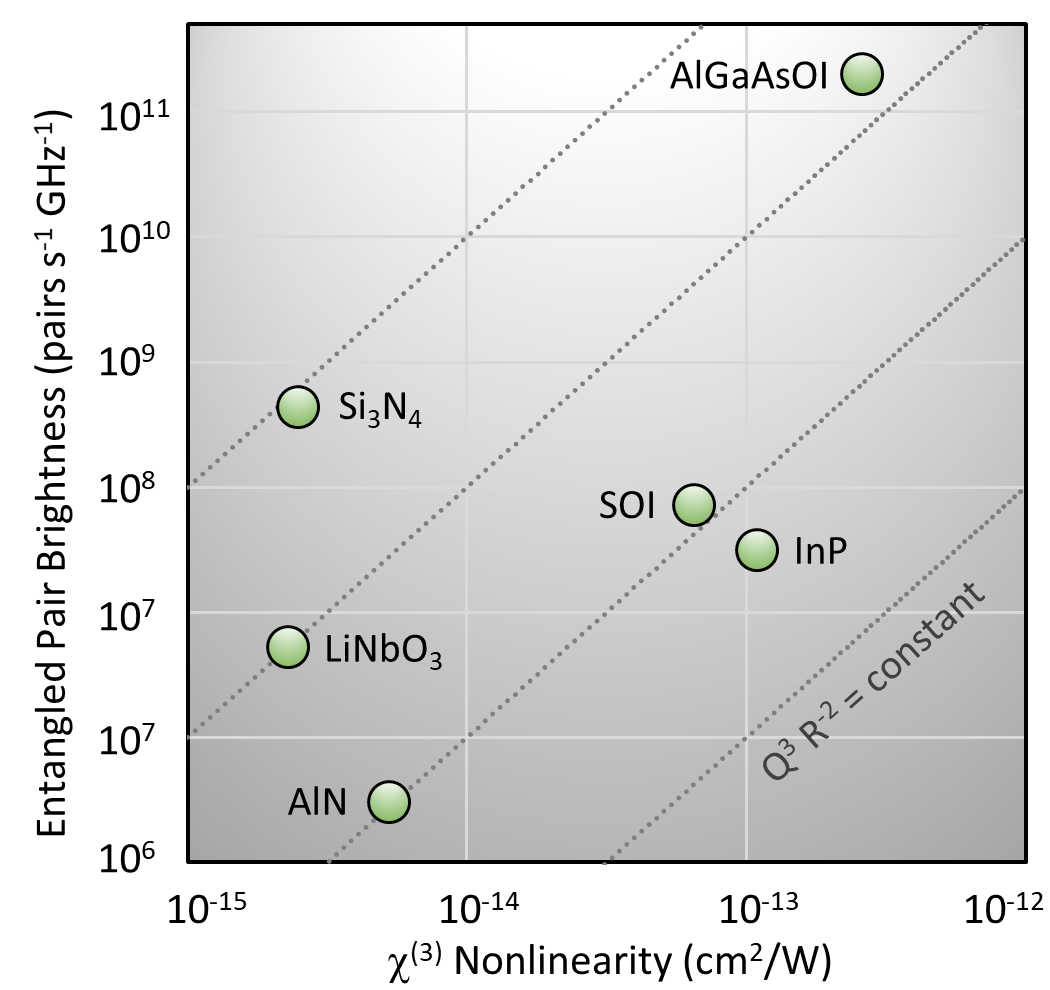}
\caption{On-chip entangled-photon pair brightness plotted as a function of the $\chi^{(3)}$ nonlinear coefficient of the material. All brightness values are normalized to 1 mW on-chip power for comparison. LiNbO$_3$ and AlN generate entangled-photon pairs via $\chi^{(2)}$ processes, but they are plotted here using their $\chi^{(3)}$ coefficients for simplicity. Isolines of constant $Q^3R^{-2}$ are illustrated by the dashed lines. The data for the materials are estimated from the references shown in Table 1.}
\label{fig:brightness}
\end{figure}

Strategies to improve on-chip entangled-photon pair generation have typically focused on improving the quality factor of microcavities while simultaneously reducing the cavity length. Exploring new material platforms with higher nonlinear coefficients has been limited by the attainable quality factor of these materials. By leveraging our recent advances in compound semiconductor nanofabrication \cite{Xie2020}, we achieve ultralow waveguide loss ($<0.4$ dB/cm) and high microring resonator quality factor ($Q>1$ million). In this work, we report the first demonstration of entangled-photon pair generation in AlGaAsOI. The high $Q$ and large third-order nonlinearity of AlGaAs lead to more than a 500-fold improvement of the on-chip pair brightness compared to all other photonic platforms as shown in Figure \ref{fig:brightness}. The photon quality also remains exceptional with a Bell-state violation measurement revealing a $97.1\pm0.6\%$ visibility, coincidence-to-accidental ratio of more than 4350 limited by the loss in our optical setup, and a heralded single-photon $g_H^{\left(2\right)}\left(0\right) = 0.004\pm0.01$. Collectively, these values yield an ultra-high quality entangled- and heralded-photon source that surpasses sources from all other integrated photonic platforms, as shown in Table 1.

\begin{table*}[!ht]
\centering
\begin{ruledtabular}
\begin{tabular}{llllllllll}

Platform & Type & Q & PGR & Brightness & CAR & Visibility & $g_H^{(2)}$ & Ref. \\
       &          &   &  [GHz]  &  [pairs s$^{-1}$ &   &   &   &  \\
       &          &   &         &   GHz$^{-1}$] &   &   &   &  \\
\hline
AlGaAsOI & SFWM & $1.2\times10^6$ & 20 & $2\times10^{11}$ & $2697\pm260$ & $97.1\pm0.6\%$ & 0.004$\pm$ 0.01 & this work \\

SOI & SFWM & $\sim 10^5$ & 0.149 & $7.1\times10^{7}$ & 532$\pm$35 & $98.9\pm 0.6\%$ & 0.0053$\pm$0.021 & \cite{Ma2017} \\

InP & SFWM & $4\times10^4$ & 0.145 & $3.1\times10^{7}$ & 277 & $78.4\pm 2\%$ & - & \cite{Kumar2019} \\

Si$_3$N$_4$ & SFWM & $2\times10^6$ & 0.004  & $4.3\times10^{8}$ & $\sim10$ & $90\pm 7\%$ & - & \cite{Ramelow2015} \\

LiNbO$_3$ & SPDC & - & 0.023 & $3\times10^{5}$ & 668$\pm$1.7 & - & - & \cite{Zhao2020} \\

AlN & SPDC & $1.1\times10^5$ & 0.006 & $5.3\times10^{6}$ & - & - & 0.088$\pm$0.004 & \cite{Guo2017} \\

\end{tabular}
\end{ruledtabular}
\caption{\label{tab:EPP_Comparison} All values are reported for microring resonators except for LiNbO$_3$, which was acquired from a linear waveguide. The CAR, visibility, and $g^{(2)}_H\left(0\right)$ shown for 1 MHz PGR. Brightness and PGR shown normalized to 1 mW on-chip pump power.}
\end{table*}
Such high performances dramatically impacts the selection of quantum sources in QPICs. In recent years, self-assembled quantum dots embedded in optical microcavities have become state-of-the-art in generating quantum light since they are capable of producing entangled-photon pairs and single photons at rates thought impossible for probabilistic nonlinear sources \cite{Wang2019}. The more than two (four) orders-of-magnitude improvement in the pair generation rate (brightness) enabled by AlGaAsOI makes probabilistic sources significantly more competitive in this category, with the additional advantages of room-temperature operation, higher quality of the produced single photons and entangled-photon pairs, and intrinsic scalability afforded by microring resonator structures.

\vspace{-15pt}
Although the AlGaAsOI platform boasts groundbreaking values in photon-pair generation, much exciting work remains to be done. Brighter entanglement sources require lower on-chip pump power for a given pair generation rate. This relaxes the requirements of on-chip filters for pump rejection. Compared to current state of the art quantum PICs based on SOI, the lower waveguide loss of AlGaAsOI can significantly reduce the overall system loss, allowing more components to be accommodated for applications requiring system-level integration, such as multi-qubit quantum computation. Another advantage is on the tuning side: AlGaAs has a factor of two larger thermo-optic cofficient than that of Si, which can enable more efficient thermo-tuners. The electro-optic and piezo-optic effects provide a novel tuning scheme for scenarios where high-speed operation or cryogenic temperatures are required. Therefore, AlGaAsOI holds exciting prospects for all-on-chip quantum photonic integrated circuits, where tunable lasers, nonlinear sources, distributed Bragg grating reflectors, Mach-Zehnder interferometers, high-speed electro-optic modulators, demultiplexers, and chip-to-fiber couplers can be monolithically integrated into the same AlGaAsOI platform without the need for complex heterogeneous integration techniques. Moreover, the large $\chi^{(2)}$ nonlinearity of AlGaAs \cite{Chang2018} can be leveraged for tuning various photonic elements at cryogenic temperatures needed for superconducting nanowire single-photon detector integration and operation.

\vspace{-15pt}
All-on-chip integration with this platform will significantly improve the performance of our devices by reducing the optical loss without sacrificing functionality as has been demonstrated on silicon-on-insulator \cite{Harris2014,Piekarek2017,Kumar:20}. The combination of our sources with low-loss, high-performing photonic components and detectors could lead to orders-of-magnitude improvement in the computation and communication rates of quantum processors, quantum transceivers, and entanglement distribution needed for quantum networks.




\begin{acknowledgements}

We thank PhotonSpot, ID Quantique, Joe Campbell from the University of Virginia, and Nadir Dagli from the University of California, Santa Barbara for technical assistance, Chenlei Li for assistance in device design, Chao Xiang for help in taking resonator photos. This work was supported by the NSF Quantum Foundry through Q-AMASE-i program Award No. DMR-1906325. G.M. acknowledges support from AFOSR YIP Award No. FA9550-20-1-0150. T.J.S., J.E.C., and L.C. contributed equally to this work.
\end{acknowledgements}





\appendix
\section{\label{app:exp}Experimental Design}

\begin{figure}[!ht]
    \centering
    \includegraphics[width=\columnwidth, height=4.6in]{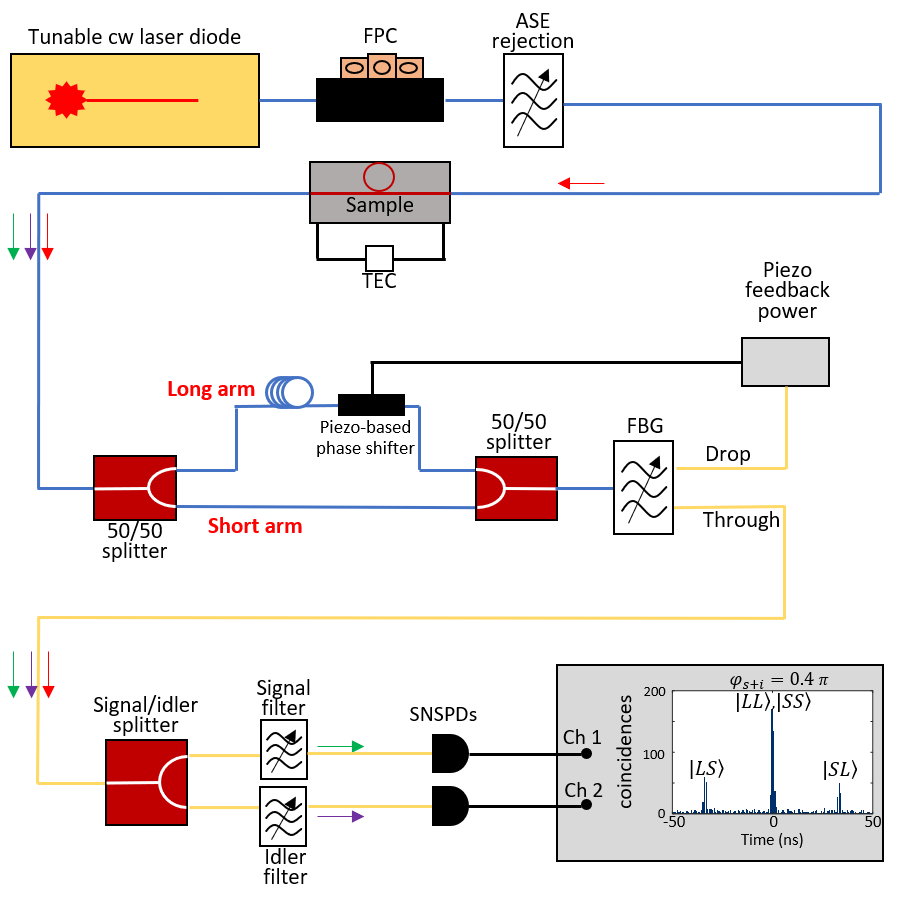}
    \caption{Schematic of the experimental setup for a fiber-based folded Franson-type interferometer. The tunable CW laser diode is swept and held at the resonance wavelength of the microring resonator. The laser is sent through ASE rejection filters and coupled via lensed fiber onto and off of the photonic chip. The light is split into a short and long arm of an interferometer. A piezo-based phase shifter is used to modify the phase of the photons that travel through the long arm. The pump photons are removed via an FBG and filters on the signal and idler channels. The signal and idler channels are coupled to SNSPDs to determine the count rates.}
    \label{fig:experiment}
\end{figure}

The full experimental design is depicted in Figure \ref{fig:experiment}. For the measurement of the singles and coincidence counts, the interferometric setup shown in Figure \ref{fig:experiment} (the short and long arm with the piezo-electric phase shifter) is bypassed. Other than this change for the singles and coincidence measurements, the rest of the experimental design remains consistent. A continuous-wave Koshin Kogaku LS601A series precision tunable laser source was stepped from approximately 0.1 nm below the resonance wavelength of the microring resonator to the resonance wavelength. The laser sweep starts below the actual resonance wavelength because the ring resonance red-shifts due to local heating of the resonator as the wavelength approaches the resonance. The laser was set to its maximum output power of 2.0 dBm (15.8 mW) and sent through a variable optical attenuator (VOA) to allow for adjustment of the input power into the chip. Etalon-based tunable fiber optic filters were placed after the VOA to provide sideband filtering of the laser. After the filters, a 99/1 fiber-based splitter was used to monitor the input power onto the chip. A lensed fiber with a numerical aperture of 0.11 was used to couple the light onto and off of the photonic chip. The coupling loss was approximately 5 dB per facet and was documented for each experiment. The lensed fiber was oriented such that the incoming light was in the transverse electric mode. The temperature of the chip was maintained using a thermo-electric cooler that was set to 20 $\degree$C. A fiber Bragg grating (FBG) was used after the chip for pump rejection with the drop channel used to monitor the power output from the chip during the experiment. The remaining light from the FBG was split using a 50/50 fiber-based beamsplitter and sent to the signal and idler filter channels. An array of four etalon-based tunable fiber optical filters were used for a total pump suppression of over 150 dB. The singles counts were monitored using superconducting nanowire single photon detectors (SNSPDs) from PhotonSpot operating at 0.77 K. Using a time-correlated single-photon counting (TCSPC) module, the signal and idler counts were recorded for 10 minutes and averaged. The scan was started with the laser set to a slightly off-resonance wavelength to determine the background counts on the detectors (from both the dark counts of the system and any pump photons that reach the detector).

After collecting the raw count data from the TCSPC module, the count rates were corrected to account for the filter losses present in each channel as well as the background counts present before the laser reached the resonance wavelength. The background counts were taken as the one minute average of the counts on each detector before the laser sweep began. To assess the loss at each filter, the laser was set to the wavelength of the signal (idler) and each of the filters on the signal (idler) channel was assessed for the loss at that wavelength. The total loss from all of the filtering and the 50/50 beamsplitters was 19.4 dB for the idler channel and 13.6 dB for the signal channel (in addition to the $\sim5$ dB facet loss). This variation is due to the different losses in the etalon-based filters. The singles counts of both the signal and the idler are fitted on a quadratic scale as shown in the main text.

To complete the coincidence measurements, the TCSPC module was set to trigger when a photon arrived on the signal channel and measure the difference in arrival time on the idler channel. The data was collected with integration times between 10 and 180 minutes dependant on the on-chip power. Larger integration times were required for lower optical powers as the coincidence-to-accidental calculation requires nonzero accidental counts. Contributors to the accidental counts include lost pairs, dark counts, and excess pump photons. Since almost zero pump photons reach the detectors at low input powers, the accidental counts are very low and require long integration times.

\section{\label{app:singlescan}Example Singles Scan}
\begin{figure}[!hbt]
    \centering
    \includegraphics[width=\columnwidth, height=3.3in]{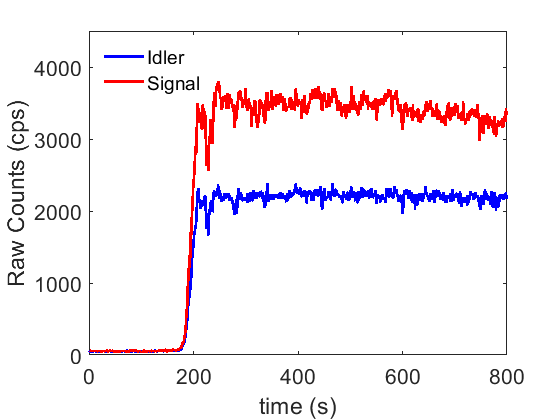}
    \caption{Example singles scan for an on-chip power of -15.14 dBm (30.6 $\mu$W). The scan begins with the laser set to an off-resonance wavelength. The laser is then swept to the resonance wavelength, reaching the resonance wavelength at approximately 200 seconds. The laser is held at this wavelength for 10 minutes to allow for the average singles rate to be determined.}
    \label{fig:Singles}
\end{figure}

To determine the singles counts for the signal and idler channels, the pump laser was swept from an off-resonance wavelength to the resonance wavelength and held on resonance for 10 minutes. An example scan for an on-chip power of -15.14 dBm (30.6 $\mu$W) is shown in Figure \ref{fig:Singles}. The initial counts represent the off-resonance counts on the detector. Ideally, these counts would be identical to the dark counts of the SNSPDs, but for larger on-chip powers, some of the pump photons still reach the detectors. Starting after approximately 200 seconds, the laser has reached the resonance wavelength and is held at this wavelength for 10 minutes. The reported singles counts are the average of this 10 minute window corrected for the background counts present when the laser is off resonance. The idler channel has lower counts due to a larger filter loss on this channel relative to the signal channel.

\section{\label{app:coinc}Example Coincidence Measurement}
\begin{figure}[!ht]
    \centering
    \includegraphics[width=\columnwidth, height=3.3in]{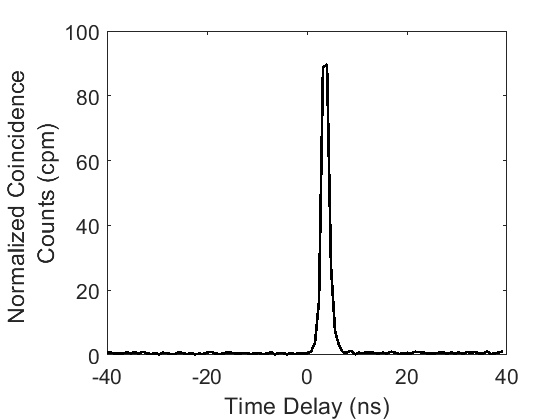}
    \caption{Example histogram showing the normalized coincidence counts (in counts per minute) as a function of the time delay. The slight offset from 0 is due to unequal filter path lengths between the signal and idler channels.}
    \label{fig:CoincidenceScan}
\end{figure}

An example histogram of a coincidence measurement at -15.71 dBm (26.9$\mu$W) is shown in Figure \ref{fig:CoincidenceScan}. The coincidence counts are integrated for 10 minutes or until the accidental counts average to a nonzero value (longer integration times for lower powers up to 180 minutes).

\clearpage
\bibliography{bibliography}

\end{document}